\documentclass[pra,showpacs,twocolumn]{revtex4}

\usepackage{dcolumn}
\usepackage{graphicx}
\usepackage{epsf}
\usepackage{amsmath}
\usepackage{bm}
\usepackage{bbm}
\usepackage{color}

\newcommand{\ee}{\mathrm e}
\newcommand{\ii}{\mathrm i}
\newcommand{\calV}{\mathcal{V}}

\begin{document}

\sloppy

%
%
\title{Separation of Transitions with Two Quantum Jumps from Cascades}

\author{Ulrich D.~Jentschura}

\affiliation{Department of Physics, Missouri University of Science
and Technology, Rolla, Missouri 65409-0640, USA}

\begin{abstract}
We consider the general scenario of an excited level $|i\rangle$ of a quantum
system that can decay via two channels: (i) via a single-quantum jump to an
intermediate, resonant level $|\overline m\rangle$, followed by a second
single-quantum jump to a final level $|f\rangle$, and (ii) via a two-quantum
transition to a final level $|f\rangle$. Cascade processes $|i\rangle \to
|\overline m\rangle \to | f\rangle$ and two-quantum transitions $|i\rangle \to
|m\rangle \to |f\rangle$ compete (in the latter case, $|m\rangle$ can be both a
nonresonant as well as a resonant level).  General expressions are derived
within second-order time-dependent perturbation theory, and the cascade
contribution is identified.  When the one-quantum decay rates of the virtual
states are included into the complex resonance energies that enter the
propagator denominator, it is found that the second-order decay rate contains
the one-quantum decay rate of the initial state as a lower-order term.
For atomic transitions, this implies
that the differential-in-energy two-photon transition rate with complex
resonance energies in the propagator denominators can be used to good accuracy 
even in the vicinity of resonance poles.
\end{abstract}


\maketitle

%
%
\section{Introduction}
\label{intro}

In this article, we consider quite a general problem which is illustrated on
the basis of the radiative decay of excited atomic levels.  Let us suppose that
an initial state $|i\rangle$ of a quantum system can decay into a final state
$|f\rangle$ via an intermediate, virtual state $|m\rangle$, under the influence
of an interaction potential $V$, with relevant matrix elements $V_{fm}$ and
$V_{mi}$.  If all available levels $|m\rangle$ are nonresonant, then the decay
rate can be computed using time-ordered second-order perturbation
theory~\cite{Sa1994Mod}. One famous example is the decay of the $2S$ state
of hydrogen, whose decay to the ground state is dipole-forbidden.
Nevertheless, the main contribution to the 
$2S$ state decays to the ground state is caused by the very 
{\em electric dipole} coupling of the bound electron to the 
quantized electromagnetic field: the transition proceeds via virtual 
$nP$ levels ($n \geq 2$), which are all nonresonant in the nonrelativistic 
approximation. One has to formulate the problem in second-order 
as opposed to first-order time-dependent perturbation theory.

Cascade decay accompanying the process
$| i \rangle \to | m \rangle \to |f \rangle$
can proceed when some of the available virtual levels $|\overline
m\rangle \in \{ | m \rangle \}$ are resonant. In that case, the atom may
first undergo a transition $|i\rangle \to |\overline m\rangle$, then
$|\overline m\rangle \to |f\rangle$ (cascade decay).  
An example is the decay $3S \to nP \to 1S$ in atomic hydrogen,
where the atom may first radiate a photon at the resonant frequency of the
$3S \to 2P$ transition, and then radiate a second photon at the 
resonant frequency of the $2P \to 1S$ transition.
However, the transition $3S \to nP \to 1S$ may also proceed
via a nonresonant $nP$ level, in which case it is a true
two-photon (two-quantum) transition. 
Indeed, the second-order transition amplitude for the sum of the 
processes $3S \to nP \to 1S$ ($n$ being summed over) 
contains both the transition amplitude due to nonresonant virtual states as 
well as the transition amplitude due to resonant intermediate 
states. Strictly speaking, the situation is even a little more 
complicated: the electric-dipole coupling of the atom 
with the radiation field couples a state with the atom in the $3S$
state to a combined atom$+$field state with the atom in the 
$|nP\rangle$ state and one photon in the radiation field.
We can denote this state as $|m \rangle = |nP, 1_{\vec{k}\lambda}\rangle$
for the particular transition mentioned. 
Here, $\vec{k}$ is the photon wave vector, and $\lambda$ is its polarization.
Unless simultaneously 
$n=2$ {\em and} the photon fulfills the resonance condition 
$E_{3S} - E_{2P} = \hbar c \, k$, where $k \equiv |\vec{k}|$
is the wave number of the photon,
the intermediate state 
$|m \rangle$ is nonresonant. An intermediate 
level $|m \rangle = |2P, 1_{\vec{k}'\lambda}\rangle$ with 
$E_{3S} - E_{2P} \neq \hbar c \, k'$ constitutes a nonresonant 
level even if the atomic part of the intermediate state---the $2P$ level---can
become resonant.  To give another example,
an intermediate state $|m \rangle = |4P, 1_{\vec{k}\lambda}\rangle$ with 
arbitrary $\vec{k}$ is always nonresonant because there
is no photon frequency available which could turn this level into a 
resonant state. The question then is how to separate the 
decay through resonant intermediate states from the decay 
via nonresonant intermediate states. Certainly, it is impossible 
to do this by excluding the $|2P\rangle$ level from the sum
over the intermediate atomic levels, because this level can be both resonant 
(if the photon frequency in the intermediate state 
is resonant with respect to the $3S \to 2P$ transition) or 
nonresonant (if the photon frequency in the intermediate state 
is nonresonant with respect to the $3S \to 2P$ transition).
The exclusion of the $2P$ state had been proposed in 
Ref.~\cite{CrTaSaCh1986} but has since been 
scrutinized~\cite{Je2008,JeSu2008}.

Related questions are investigated here in more general terms: How 
can we formulate the problem, within time-dependent second-order 
perturbation theory, so that the resonant intermediate levels in the process 
$| i \rangle \to |m \rangle \to | f \rangle$ are separated from the  nonresonant levels,
and so that the cascade contribution due to resonant intermediate 
levels $| \overline m \rangle$ is clearly identified within the 
time-dependent formalism? In order to answer this question,
we first recall that under rather general assumptions about the 
process, the intermediate
states $|m\rangle$ represent a {\em continuum} of states. 
This is the case even
in transitions of {\em discrete} atomic levels because the intermediate states
$|m\rangle$ in this case are product states of the atom in a discrete state and
one or more excited modes of the electromagnetic field. 
While the bound states
of the atom are discrete, the photon modes represent a continuum of energies.
In particular, the photon wave vector $\vec{k}$ represents a continuous variable.
A resonant process involves a transition to a lower atomic level with a
simultaneous emission of a photon of the resonant frequency; in that case, the
resonant state $|\overline m\rangle$ is an eigenstate of the unperturbed
Hamiltonian of atom$+$radiation field with exactly the same energy as the
initial state (the sum of the energies of the lower atomic state and of the
energy of the radiated photon is equal to the energy of the initial atomic
state). 

When the decay $|i\rangle \to |m\rangle \to |f\rangle$ can proceed
via a {\em resonant} state $|\overline m\rangle$ which can
be reached from $|i\rangle$ via a single quantum jump, we have to take into
account both possibilities: (a)~the
sequential transition (cascade) and (b)~the two-quantum transition via the 
nonresonant levels. One possibility to identify the cascade
within time-dependent perturbation theory is given by the 
functional form of its time dependence:
for a cascade decay $|i \rangle \to |\overline m\rangle \to |f\rangle$, 
the probability of finding the system in the final state
$|f\rangle$ is proportional to the square of the elapsed time $t$:
The system first has to make a quantum jump $|i\rangle \to |\overline m\rangle$,
leading to a linear increase (with time) of the 
population of the resonant intermediate level $|\overline m\rangle$.
The second quantum jump $|\overline m\rangle \to |f\rangle$ 
then leads to a quadratic increase of the probability of finding 
the system in state $|f\rangle$ with time.
By contrast, the true nonsequential two-quantum transition 
$|i\rangle \to |m\rangle \to |f\rangle$ via nonresonant
intermediate states leads to 
a linear increase (with time) of the probability of finding 
the system in state $|f\rangle$ with time.
Here, we identify, in a general formalism, those contributions of the two-quantum
transition which contribute to the linear behavior (in time), and separate
them from the (quadratic in time) cascade effect. 

We follow Ref.~\cite{Sa1994Mod} in our conventions
and proceed as follows: First, the basics of a single-quantum
transition are recalled (Sec.~\ref{sec2}).
We then proceed to the discussion of a two-quantum transition
without cascades (Sec.~\ref{sec3}), before including the 
cascades/resonant levels in Sec.~\ref{sec4}.
Conclusions are reserved for Sec.~\ref{sec5}.
The interaction is switched off adiabatically
in the distant past and in the distant future,
but the rate is calculated near $t=0$.
We work in natural units ($\hbar = c = \epsilon_0 = 1$).

%
%
\section{SINGLE--QUANTUM TRANSITION}
\label{sec2}

%
%
\subsection{General formulation}
\label{sec2A}

Following Chap.~5 of Ref.~\cite{Sa1994Mod}, we first consider a single-quantum
transition $|i\rangle$ to $|f\rangle$, with $c_f(t)$ being the time-dependent
expansion coefficient of the final-state Hilbert vector with respect to the
state $|f\rangle$. The interaction is adiabatically damped on in the infinite
past $t \to - \infty$ and suppressed by an exponential factor $\exp( \eta t)$,
with $\eta > 0$ being an infinitesimal parameter.  We then start the time
evolution with $c_f(0) = 0$ and $c_i(0) = 1$ (initially, the system is in the
state $|i\rangle$).  For the complex 
probability amplitude $c_f(t)$ of finding the 
system in state $|f\rangle$ at time $t$, one finds
[see Eq.~(5.8.2) of Ref.~\cite{Sa1994Mod}],
\begin{equation}
\label{cf}
c_f(t) = - \ii \int\limits_{-\infty}^{t} 
V_{fi} \, 
{\rm e}^{\eta \, t'} \,
\ee^{\ii \omega_{fi} t'} \, 
{\rm d}t'
= -\frac{ {\rm e}^{\eta t + {\ii} \omega_{fi} t} }
{\omega_{fi} - {\ii} \eta} \,  V_{fi},
\end{equation}
where $V_{fi}$ is the matrix element of the interaction Hamiltonian
$V$ in the Schr\"{o}dinger picture,
i.e.,~$V_{fi} = \langle f | V | i \rangle$.
Note that there is a somewhat subtle difference
between the interaction Hamiltonian $V$ in the Schr\"{o}dinger picture,
and the interaction Hamiltonian 
$\exp(\ii H_0 t) V \exp(-\ii H_0 t)$ in the interaction picture,
because in the latter case, matrix elements of $V$ acquire a time dependence.
This time dependence is explicitly written out in the term 
$V_{fi} \, \ee^{\ii \omega_{fi} t'}$ in Eq.~\eqref{cf}.

In the case of an electric-dipole transition in an atom,
$V$ is the coupling of the bound electron to the 
quantized radiation field.
The expression $\omega_{fi} = E_f - E_i$ is the 
energy difference of the initial and final state 
of the transition
with respect to the unperturbed Hamiltonian $H_0$
of the system.
In the case of an electric-dipole transition in an atom,
$H_0$ is the sum of the unperturbed Hamiltonian of the 
atom and of the electromagnetic Hamiltonian counting the
modes of the radiation field.  From Eq.~\eqref{cf}, we find
$| c_f(t) |^2 =
{\rm e}^{2 \eta t} \, | V_{fi} |^2 /
(\omega^2_{fi} + \eta^2)$.
Differentiating this expression with respect to time,
we obtain
\begin{align}
\frac{\rm d}{{\rm d}t} | c_f(t) |^2 =& \
\frac{ {\rm e}^{2 \eta t} \, 2 \, \eta \, | V_{fi} |^2 }
{\omega^2_{fi} + \eta^2} \,.
\end{align}
With the identification 
[see Eq.~(5.8.5) of Ref.~\cite{Sa1994Mod}]
\begin{equation}
\label{todelta}
\frac{\eta}{\omega^2_{fi} + \eta^2} \to 
\pi \, \delta( \omega_{fi} ) \,, \qquad
\eta \to 0^+ \,,
\end{equation}
we obtain in the limit $\eta \to 0^+$ 
\begin{align}
\label{FGR}
\Gamma^{(1)}_{fi} = 
\left. \frac{\rm d}{{\rm d}t} | c_f(t) |^2 \right|_{t=0} =
2 \, \pi \, | V_{fi} |^2 \delta( \omega_{fi} ) \,,
\end{align}
where by definition,
$\Gamma^{(1)}_{fi}$ is the decay rate associated
with the transition $|i\rangle \to |f\rangle$ via a single 
quantum jump (we reemphasize that the time derivative is taken at $t=0$). 
This result is known as Fermi's golden rule
[see Eqs.~(5.6.35) and~(5.8.6) of \cite{Sa1994Mod}].

One might wonder why the Dirac $\delta$ persists
in the final result, although Fermi's Golden Rule 
is known to be directly applicable to 
experimentally relevant calculations,
and an expression containing a Dirac $\delta$ might otherwise 
be assumed not to be 
applicable to an experiment.
Just after Eq.~(5.6.35) of Ref.~\cite{Sa1994Mod},
which is equivalent to Eq.~\eqref{FGR} in this work,
it is stated that the final state must be integrated
over an (infinitesimal) interval about the
final-state energy. This statement is useful, but it 
may need a more complex explanation for a full elucidation.
Indeed, the solution to this question involves two observations:
(a) that Eq.~\eqref{FGR} needs to be summed over the 
state variables of the radiated quanta (in the case of an 
atomic transition, photons) in order to 
make experimentally relevant predictions,
and (b) that the Dirac $\delta$
disappears when all possible energies and
all possible polarizations of the emitted
quanta are taken into account in the final state.
In order to illustrate this aspect, we now 
discuss the application of Eq.~\eqref{FGR} to an electric 
dipole transition in an atom.

%
%
\subsection{Specialization to an atomic transition}
\label{sec2B}

In the case of an electric-dipole transition of an atom, 
the final state $|f\rangle$
is a product state of the atom in state $|f_A\rangle$ 
and one radiated photon $|1_{\vec{k}\lambda} \rangle$ 
in the radiation field. 
In the following, we will write a 
general product state $|f\rangle$ 
of the system composed of the atom$+$radiation field as
\begin{equation}
| f \rangle = | f_A, \tilde f \rangle
\end{equation}
where $| f_A \rangle$ is atomic part of the product state,
and $| \tilde f \rangle$ is the photon part of the 
product state. 
The unperturbed Hamiltonian of the system is 
\begin{equation}
\label{H0}
H_0 = \sum_{f_A} E_{f_A} | f_A \rangle \, \langle f_A | +
\sum_{\vec{k}\lambda} 
k \, a^+_{\vec{k}\lambda} \, a_{\vec{k}\lambda} \,,
\end{equation}
where the $a_{\vec{k}\lambda}$ and $a^+_{\vec{k}\lambda}$ 
are photon annihilation and creation operators
(here, we work in a representation with a finite 
normalization volume $\calV$, i.e., $[ a_{\vec{k}\lambda}, 
a^+_{\vec{k}'\lambda'}] = \delta_{\vec{k} \vec{k}'} \delta_{\lambda\lambda'}$).
Eigenstates of the Hamiltonian~\eqref{H0} 
are product states of the atom in 
eigenstate $|f_A\rangle$
and a Fock state of the electromagnetic field 
such as $| 1_{\vec{k}\lambda} \rangle$.

In the Schr\"{o}dinger picture,
the dipole interaction of an electron at point $\vec{x}$ with the
quantized electromagnetic field is given by
\begin{align}
\label{Hint}
V =& \; -e \, \vec{x} \cdot \vec{E} \,,
\\
\vec{E} =& \; 
\, \sum_{\vec{k}\lambda} 
\sqrt{\frac{k}{2 \calV}}
\left( 
\hat{\epsilon}_{\vec{k}\lambda} \, a_{\vec{k}\lambda} +
\hat{\epsilon}_{\vec{k}\lambda} \, a^+_{\vec{k}\lambda} 
\right) \,,
\nonumber
\end{align}
with the unit polarization vectors $\hat{\epsilon}_{\vec{k}\lambda}$
and the electric field operator $\vec{E}$.
In atomic physics, one distinguishes between the
($\vec{p}\cdot \vec{A}$) and
($\vec{x}\cdot \vec{E}$) forms of the interaction with the electromagnetic
field. The former is called the velocity gauge because of the 
appearance of the electron momentum in the interaction Hamiltonian.
The latter is commonly referred to as the length gauge,
because the electron coordinate $\vec{x}$ in the interaction Hamiltonian
has physical dimension of length.
In some situations, the length gauge is preferable because
the interaction is formulated in terms
of physically observable electric field strength $\vec{E}$ instead of the
gauge-dependent vector potential $\vec{A}$
(see Refs.~\cite{Ko1978prl,Ko1983,ScBeBeSc1984}). All results
presented here are given in the length gauge.

While the initial state of the atom $|i_A\rangle$ 
and the final state of the atom $|f_A\rangle$ are
well-defined for an atomic decay rate,
we have to sum over the degrees of freedom of the radiated photons
in order to obtain the decay rate for the one-photon
transition $|i_A\rangle \to |f_A\rangle$.
The Dirac $\delta$ function in Eq.~\eqref{FGR}
ensures the fulfillment of the resonance condition.
For a one-photon final state, we
can replace
\begin{equation}
\sum_{\tilde f} \to
\sum_{\vec{k}\lambda} 
\end{equation}
for the sum over the photon degrees of freedom of the final state.
Indeed, the atomic one-photon ($1\gamma$) decay rate for the
transition $|i_A\rangle \to |f_A\rangle$, which we denote
as $\Gamma^{(1\gamma)}_{f_A \, i_A}$, is obtained as the sum
\begin{align}
\label{one-gamma}
\Gamma^{(1\gamma)}_{f_A \, i_A} = & \; 
\sum_{\tilde f}
\Gamma^{(1)}_{fi} =
\sum_{\vec{k}\lambda} 
\Gamma^{(1)}_{fi} =
\sum_{\vec{k}\lambda} 
\Gamma^{(1)}_{|f_A, 1_{\vec{k}\lambda}\rangle, \, |i_A, 0 \rangle} 
\\
=& \; 2 \pi \sum_{\vec{k}\lambda} 
\delta(E_{f_A} - E_{i_A} - k)
\left| \left< f_A, 1_{\vec{k}\lambda} \left| V
\right| i_A, 0 \right> \right|^2 \,,
\nonumber
\end{align}
where we recall that the sum over 
$\vec{k}$ and $\lambda$ transforms into an integral 
in the continuum limit,
\begin{equation}
\label{continuum}
\sum_{\vec{k}\lambda} \to
\calV \, \sum_{\lambda} \int \frac{{\rm d}^3 k}{(2\pi)^3} \,.
\end{equation}
This integral cancels the Dirac $\delta$. 
We reemphasize that the Dirac $\delta$ function is 
eliminated after a  summation over specific 
degrees of freedom of the final state, namely, the 
degrees of freedom of the electromagnetic field.

One might wonder why the single-quantum transitions apparently 
conserve energy according to the above formalism
[persistence of the $\delta(\omega_{fi})$ in Eq.~\eqref{FGR}], while 
spontaneous decay of an atomic state always tends to lower
the energy of the bound electron. The answer is that the 
final state of the process, which is a bound electron in a lower
state plus a single resonant photon, has the same energy 
as the initial state (electron in the excited state and no 
photon in the radiation field). This is manifest in the 
expression $\delta(\omega_{fi}) = \delta(E_{f_A} - E_{i_A} - k)$ in 
Eq.~\eqref{one-gamma}.

In view of Eq.~\eqref{Hint}, the transition matrix element 
$\left< f_A, 1_{\vec{k}\lambda} \left| V \right| i_A, 0 \right>$
in Eq.~\eqref{one-gamma} can be written as
\begin{equation}
\left< f_A, 1_{\vec{k}\lambda} \left| 
\left( -e \, \vec{x} \cdot \vec{E} \right) \right| i_A, 0 \right>
= -e \, \sqrt{\frac{k}{2 \calV}} \,
\hat{\epsilon}_{\vec{k}\lambda} \cdot
\left< f_A \left| \vec{x} \right| i_A \right> \,,
\end{equation}
where we denote the atomic component of the 
bra and ket vectors by a subscript $A$.
The sum over the photon modes in Eq.~\eqref{Hint} collapses
because there is exactly one definite photon mode 
occupied in the state $| f_A, 1_{\vec{k}\lambda} \rangle$.
Summing over the available photon modes in the 
exit channel, we obtain ($k \equiv |\vec{k}|$)
\begin{align}
\label{manif}
& \Gamma^{(1\gamma)}_{f_A \, i_A} = \; 
\sum_{\vec{k}\lambda} 
2 \, \pi \, | V_{fi} |^2 \delta( \omega_{fi} ) 
\nonumber\\
&= \; 
\sum_{\vec{k}\lambda} 
2 \, \pi \, e^2 \, 
\frac{k}{2 \calV} \,
\left| \hat{\epsilon}_{\vec{k}\lambda} \cdot
\left< f_A \left| \vec{x} \right| i_A \right> \right|^2 \,\, 
\delta( E_{f_A} - E_{i_A} - \omega_{\vec{k}} ) 
\nonumber\\
&= \; 
\sum_\lambda
\int \frac{{\rm d}^3 k}{(2\pi)^3} \,
4 \pi^2 \alpha \, k \,
\left| \hat{\epsilon}_{\vec{k}\lambda} \cdot
\left< f_A \left| \vec{x} \right| i_A \right> \right|^2 
\nonumber\\
& \qquad \times \delta( E_{f_A} - E_{i_A} - \omega_{\vec{k}} ) 
\nonumber\\
&= \; 
\int \frac{{\rm d} \Omega_k}{4\pi} 
2 \alpha (E_{f_A} - E_{i_A})^3 
\delta^{{\rm T},jk} 
\left< f_A \left| x^j \right| i_A \right>
\left< i_A \left| x^k \right| f_A \right>
\nonumber\\
&= \; 
\frac{4}{3} \alpha \, (E_f - E_i)^3 \,
\left| \left< f_A \left| \vec{x} \right| i_A \right> \right|^2 \,,
\end{align}
which is the familiar result for a one-photon electric-dipole
decay rate. The transverse delta function is
$\delta^{{\rm T},ij} =
\delta^{ij} - k^i \, k^j/k^2$. We denote the 
Cartesian components of a vector by superscripts.
Note, in particular, that the sum over the photon modes in 
Eq.~\eqref{Hint} is not enough in order to 
calculate the familiar expression for the one-photon
decay rate; an additional summation over final states
is necessary.

%
%
\section{TWO--QUANTUm TRANSITION WITHOUT CASCADES}
\label{sec3}

%
%
\subsection{General Formulation}
\label{sec3A}

In second-order time-dependent perturbation theory, 
the amplitude $c_f(t)$ to find the system in 
state $|f\rangle$ at time $t$ due to the 
transition $|i\rangle \to |m\rangle \to |f\rangle$ is given by
\begin{equation}
\label{cfnocascades}
c_f(t) = (-{\ii})^2 \!\!\!\!
\int\limits_{-\infty}^t {\rm d}t' {\rm e}^{\eta t' + 
{\ii} \omega_{fm} t'} V_{fm}
\!\!
\int\limits_{-\infty}^{t'} {\rm d}t'' {\rm e}^{\eta t'' + 
{\ii} \omega_{mi} t''} V_{mi} \,,
\end{equation}
which leads to 
[see Eq.~(5.6.37) of Ref.~\cite{Sa1994Mod}],
\begin{equation}
\label{cf2}
|c_f(t)|^2 =
\frac{{\rm e}^{4 \eta t}}{(4 \eta^2 + \omega_{fi}^2)}
\left| \sum_m  \frac{V_{fm} \, V_{mi}}%
{\omega_{mi} - {\ii} \eta } \right|^2  \,.
\end{equation}
This is a generalization of Eq.~\eqref{cf} to second order.  When no cascades
are allowed, we can differentiate with respect to time and assume that
$\omega_{mi} \neq 0$ is always nonvanishing. In oder to fix ideas by comparison
to a concrete example, we recall that in the case of the $2S \to 1S$ two-photon
transition in atomic hydrogen, the intermediate state $|m\rangle = |nP,
1_{\vec{k}\lambda}\rangle$ has a higher energy than the initial state
$|i\rangle = |2S, 0\rangle$ where the atom is in the $2S$ state and the
electromagnetic field is in the vacuum state $|0\rangle$.  No cascades are
relevant in this case, and Eq.~\eqref{cf2} is immediately applicable.

We can thus differentiate Eq.~\eqref{cfnocascades} with respect to time and
obtain
\begin{align}
\label{Gamma2without}
\Gamma^{(2)}_{fi} =& \;  \left.
\left( \frac{{\rm d}}{{\rm d} t} |c_f|^2 \right) \right|_{t=0} 
\nonumber\\
=& \; \frac{4 \eta}{(2 \eta)^2 + \omega_{fi}^2}
\left| \sum_m  \frac{V_{fm} \, V_{mi}}%
{\omega_{mi} - {\ii} \eta} \right|^2
\nonumber\\
= & \; 
2 \pi \, \delta(\omega_{fi}) \,
\left| {\sum_m}  \frac{V_{fm} \, V_{mi}}%
{\omega_{mi}} \right|^2 \,, \qquad 
\eta \to 0^+ \,.
\end{align}
In analogy to the single-quantum transition described by
Eq.~\eqref{FGR}, the Dirac $\delta$ disappears when the final states are summed
over the experimentally relevant degrees of the radiated quanta. 
We now verify that Eq.~\eqref{Gamma2without} exactly reproduces the known
expressions~\cite{GM1931,Je2004rad} for two-photon decay rates in atoms.

%
%
\subsection{Specialization to an atomic transition}
\label{sec3B}

For a two-photon transition in an atom, we can write the initial state as
$| i \rangle = | i_A, 0 \rangle$,
where $| i_A \rangle $ is the atomic final state, 
and $|0 \rangle$ is the vacuum state of the electromagnetic field.
The intermediate state is 
$| m \rangle = | m_A, 1_{\vec{k}\lambda} \rangle$,
where the atom is in state $|m_A\rangle$, and the 
electromagnetic field is in the one-photon Fock 
state $| 1_{\vec{k}\lambda} \rangle$.
The final state is
$| f \rangle = | f_A, 1_{\vec{k}_1\lambda_1}, 1_{\vec{k}_2\lambda_2} \rangle$,
where the $|\vec{k}_i|$ and $\lambda_i$ are the wave vectors and 
polarizations of the two radiated
photons ($i = 1,2$). 

In specializing Eq.~\eqref{Gamma2without} to a two-photon transition in atoms,
we have to take into account a subtlety, which we 
outline in greater detail because it becomes relevant for all discussions in the
following. Namely,
the atomic decay rate is obtained after summing the rate
$\Gamma^{(2)}_{fi}$ over the degrees of freedom of all possible radiated photons. 
Now, if we sum the final states over all 
$\vec{k}_1 \lambda_1$ and all $\vec{k}_2 \lambda_2$, we count the 
photons twice, because the Fock state 
$| 1_{\vec{k}_2\lambda_2} , 1_{\vec{k}_1\lambda_1} \rangle$ 
obtained under the 
simultaneous exchange $\vec{k}_1 \leftrightarrow \vec{k}_2$
and $\lambda_1 \leftrightarrow \lambda_2$ is identical to the 
original state $| 1_{\vec{k}_1\lambda_1}, 1_{\vec{k}_2\lambda_2} \rangle$.
Hence,
\begin{align}
\label{two-gamma}
\Gamma^{(2\gamma)}_{f_A \, i_A} = & \; 
\frac12 \sum_{\vec{k}_1\lambda_1} \sum_{\vec{k}_2\lambda_2} 
\Gamma^{(2)}_{fi} \,.
\end{align}
The factor $1/2$ is discussed after
Eq.~(5.108) on p.~169 of the quantum field theory 
textbook~\cite{PeSc1995}
and in the text preceding Eq.~(3.316) of the textbook~\cite{GrRe1992}.

With reference to Eq.~\eqref{Gamma2without},
we now turn our attention to the 
two-quantum decay rate (without cascades).
Here, two quantum paths are possible which 
must be added coherently.
These correspond to a different time ordering for the 
emissions of the photons with photon wave vector $\vec{k}_i$
and polarization $\lambda_i$ ($i=1,2$).
Summing over the final-state photon polarizations, the result
then is 
\begin{align}
\label{manif2}
& \Gamma^{(2\gamma)}_{f_A i_A} = \; 
\frac12 \sum_{\vec{k}_1\lambda_1} 
\sum_{\vec{k}_2\lambda_2} 
2 \pi \, \delta(\omega_{fi}) \,
\left| {\sum_m}  \frac{V_{fm} \, V_{mi}}{\omega_{mi}} \right|^2 \,,
\nonumber\\
&= \; 
\sum_{\vec{k}_1\lambda_1} 
\sum_{\vec{k}_2\lambda_2} 
\pi \, e^4 \, 
\frac{k_1}{2 \calV} \,
\frac{k_2}{2 \calV} \,
\delta( E_f - E_i - k_1 - k_2 ) 
\nonumber\\
& \quad \times
\left| 
\sum_{m_A} \left( 
\frac{ \left( \hat{\epsilon}_{\vec{k}_1\lambda_1} \cdot 
\left< f_A \left| \vec{x} \right| m_A \right> \right) 
\left( \hat{\epsilon}_{\vec{k}_2\lambda_2} \cdot 
\left< m_A \left| \vec{x} \right| i_A \right> \right) }%
{E_{m_A} - E_{i_A} + k_2} \right.  \right.
\nonumber\\
& \qquad + \left. \left.
\frac{ \left( \hat{\epsilon}_{\vec{k}_2\lambda_2} \cdot 
\left< f_A \left| \vec{x} \right| m_A \right> \right)
\left( \hat{\epsilon}_{\vec{k}_1\lambda_1} \cdot 
\left< m_A \left| \vec{x} \right| i_A \right> \right) }%
{E_{m_A} - E_{i_A} + k_1} 
\right) \right|^2 \,.
\end{align}
Separating angular and radial variables for the photon energies,
we finally obtain the following known result~\cite{GM1931}
in the continuum limit [see Eq.~\eqref{continuum}]:
\begin{align}
\label{known}
& \Gamma^{(2\gamma)}_{f_A i_A}
= \; 
\frac{4 \, \alpha^2}{27 \, \pi} \,
\int\limits_0^{E_{f_A} - E_{i_A}} {\rm d} k \, k^3 \, 
(E_{f_A} - E_{i_A} - k)^3 \,
\nonumber\\
& \times
\left| 
\sum_{m_A} \left( 
\frac{ 
\left< f_A \left| x^j \right| m_A \right>
\left< m_A \left| x^j \right| i_A \right> }%
{E_{m_A} - E_{i_A} + k} \right. \right.
\nonumber\\
& \qquad \left. \left. +
\frac{ \left< f_A \left| x^j \right| m_A \right>
\left< m_A \left| x^j \right| i_A \right> }%
{E_{m_A} - E_{f_A} - k}  \right)
\right|^2  \,.
\end{align}
The integration over $k$ extends over the allowed frequency 
range for a two-photon transition~\cite{Je2009}.
The subtlety with respect to the counting of photon
modes illustrates that Eq.~\eqref{Gamma2without}
cannot be applied to atomic transitions without a proper 
interpretation of all physical quantities involved.

%
%
\section{TWO--QUANTUM TRANSITION WITH CASCADES}
\label{sec4}

%
%
\subsection{General formulation}
\label{sec4A}

We return once more to Eq.~\eqref{cf2}
which gives the result for the 
two-photon decay rate
[see also Eq.~(5.6.37) of Ref.~\cite{Sa1994Mod}],
\begin{equation}
\label{cf2repeat}
|c_f(t)|^2 =
\frac{{\rm e}^{4 \eta t}}{(4 \eta^2 + \omega_{fi}^2)}
\left| \sum_m  \frac{V_{fm} \, V_{mi}}%
{\omega_{mi} - {\ii} \eta } \right|^2  \,.
\end{equation}
In the text directly following
Eq.~(5.6.37) of Ref.~\cite{Sa1994Mod},
it is stated that the best way to deal with
the situation of a resonant intermediate state
with $\omega_{mi} \approx 0$
is to use an adiabatic turn-on
of the perturbation that leads to the 
transition. We have already incorporated this 
adiabatic turn-on into 
Eq.~\eqref{cfnocascades}. It is also stated
in Eq.~(5.6.38) of Ref.~\cite{Sa1994Mod}
that the turn-on amounts to the replacement
\begin{equation}
\omega_{mi} \to
{\omega_{mi} - {\ii} \eta } 
\end{equation}
in the denominator of the expression on the right-hand side 
of Eq.~\eqref{cf2repeat}.
Again, we have already incorporated the infinitesimal imaginary part
in Eq.~\eqref{cf2repeat}.
Here, we extend the discussion beyond that in
Ref.~\cite{Sa1994Mod} and analyze the resonant and nonresonant
levels separately.

We now assume that some of the intermediate states
of the system are close in energy to the initial
state of the process, i.e.~that there 
exist states $|\overline m \rangle$ with 
$E_{\overline m} = E_i$. We recall that 
$E_{\overline m}$ here represents the total
energy if the system. In the case of an atomic 
transition, this would be the sum of the energy of the 
intermediate atomic level and of energy of the photons
radiated.
In order to analyze this process, we restrict, in Eq.~\eqref{cf2repeat}, the sum 
over intermediate states to the resonant states $\overline m$.
Then,
\begin{equation}
\label{cf2mbar}
|\overline c_f(t)|^2 =
\frac{{\rm e}^{4 \eta t}}{(2 \eta)^2 + \omega_{fi}^2}
\left| \sum_{\overline m}  \frac{V_{f\overline m} \, V_{\overline mi}}%
{\omega_{\overline mi} - {\ii} \eta } \right|^2  \,,
\end{equation}
where $\omega_{\overline mi}$ tends to zero.
The cascade contribution associated with the 
resonant levels $| \overline m \rangle$
needs to be differentiated twice with respect to the time.
We obtain
\begin{align}
\label{expected}
& \left. \left( \frac{{\rm d}^2}{{\rm d} t^2} 
|\overline c_f|^2 \right) \right|_{t=0} =
\frac{16 \, \eta^2 }{(2 \eta)^2 + \omega_{fi}^2}
\left| \sum_{\overline m}  \frac{V_{f \overline m} \, V_{\overline mi}}%
{ \omega_{\overline mi} - {\ii} \eta } \right|^2
\nonumber\\
& \mathop{=}^{\eta \to 0^+} 
4 \pi^2 \, 
\sum_{\overline m} 
\left| V_{f\overline m} \, V_{\overline mi} \right|^2 \,
\delta(\omega_{f\overline m}) \, \delta(\omega_{fi}) 
\nonumber\\
& =
4 \pi^2 \, \sum_{\overline m} 
\left| V_{f\overline m} \, V_{\overline mi} \right|^2 \,
\delta(\omega_{f\overline m}) \, \delta(\omega_{\overline m i})  
\nonumber\\
& = \; \sum_{\overline m} \Gamma^{(1)}_{f\overline m} \, 
\Gamma^{(1)}_{\overline mi} \equiv C_{fi} \,.
\end{align}
In the last step, we define the expression $C_{fi}$ as the relevant 
cascade term which we evaluate for atomic transitions
in Sec.~\ref{sec4B} below.
We also assume that interference terms among the different
resonant levels $|\overline m\rangle$ vanish. 

Equation~\eqref{expected} is just the expected result: the level $|i\rangle$ 
feeds the resonant intermediate levels $|\overline m\rangle$
with a time dependence $\Gamma^{(1)}_{\overline mi} \, t$, 
and the resonant intermediate levels, in turn, feed the 
final state population as 
\begin{equation}
\label{cascade}
|\overline c_f(t)|^2 =
\sum_{\overline m} 
\int\limits_0^t {\rm d}t' \, \Gamma^{(1)}_{f{\overline m}} \, 
\Gamma^{(1)}_{\overline mi} \, t' =
\tfrac{1}{2} \, 
\sum_{\overline m} 
\Gamma^{(1)}_{f{\overline m}} \, 
\Gamma^{(1)}_{\overline mi} \, t^2 \,.
\end{equation}
The necessity of the 
sum over $\overline m$ is also clear, because all intermediate 
resonant levels have to be included.

Now that we have treated the resonant levels separately, we have
to subtract them from the remaining expression. We thereby
obtain a modified 
probability $\overline{|c_f(t)|^2}$ 
of finding the system in state $|f\rangle$,
\begin{align}
\label{eq2}
\overline{|c_f(t)|^2} =& \; |c_f(t)|^2 - |\overline c_f(t)|^2
\nonumber\\
=& \; \frac{{\rm e}^{4 \eta t}}{(4 \eta^2 + \omega_{fi}^2)}
\left| \sum_{m}  \frac{V_{f m} \, V_{mi}}%
{\omega_{mi} - {\ii} \eta } \right|^2 
\nonumber\\
& \; -\frac{{\rm e}^{4 \eta t}}{(4 \eta^2 + \omega_{fi}^2)}
\left| \sum_{\overline m}  \frac{V_{f\overline m} \, V_{\overline mi}}%
{\omega_{\overline mi} - {\ii} \eta } \right|^2  \,.
\end{align}
One might think that the subtraction term
(second term on the right-hand side of the above
equation) would imply, e.g., the 
subtraction of the intermediate $2P$ state in 
the two-photon decay of the 
$3S$ state of hydrogen. However, that is not the case.
The intermediate states are  quantum states of the 
coupled system of atom$+$radiation field.
As already outlined in Sec.~\ref{intro},
the product state composed of the $2P$ level
and a resonant photon would qualify as a resonant
state $|\overline m\rangle$, but a $2P$ state with a 
slightly off-resonant photon would not constitute a 
resonant intermediate state. Therefore, the $2P$ state may 
not be taken out of the 
sum over the atomic-state components of the virtual states. 
The time derivative of the subtracted 
expression $\overline{|c_f|^2}$ is
\begin{align}
\label{eq3}
\left. \left( \frac{\rm d}{{\rm d} t} 
\overline{|c_f|^2} \right) \right|_{t=0} =& \;
\frac{4 \eta }{(4 \eta^2 + \omega_{fi}^2)}
\left| \sum_{m}  \frac{V_{fm} \, V_{mi}}%
{\omega_{mi} - {\ii} \eta } \right|^2 
\nonumber\\
& \; -\frac{4 \eta }{(4 \eta^2 + \omega_{fi}^2)}
\left| \sum_{\overline m}  \frac{V_{f\overline m} \, V_{\overline mi}}%
{\omega_{\overline mi} - {\ii} \eta } \right|^2  \,.
\end{align}
The second term on 
the right-hand side of Eq.~\eqref{eq3} is divergent 
in the limit $\eta \to 0^+$ and $\omega_{\overline mi} \to 0$.
We cannot proceed without giving a physical 
interpretation to the adiabatic parameter $\eta$.
First, in the subtraction term
\begin{align}
\label{defS}
S = -\frac{4 \eta }{(2 \eta)^2 + \omega_{fi}^2}
\left| \sum_{\overline m}  \frac{V_{f\overline m} \, V_{\overline mi}}%
{\omega_{\overline mi} - {\ii} \eta } \right|^2  \,,
\end{align}
we carry out the limit $\eta \to 0^+$ 
in the prefactor; this leads to a Dirac $\delta$.
Then, for the sum over $\overline m$, we match the 
adiabatic parameter with the imaginary part of the 
interaction Hamiltonian.
The in and out states $|i\rangle$ and 
$|f\rangle$ are assumed to be asymptotic,
stable states in the infinite past and future
within the context of adiabatic perturbation theory.
Adiabatically, we therefore switch on only the 
virtual intermediate states.  We should thus replace 
\begin{equation}
\label{ident}
\eta \to \tfrac{1}{2} \Gamma^{(1)}_{fm}
\end{equation}
for every term in the sum. We then obtain
\begin{align}
\label{S}
S =& \; -2 \pi \, \delta(\omega_{fi}) \,
\left| \sum_{\overline m}  \frac{V_{f\overline m} \, V_{\overline mi}}%
{\omega_{\overline mi} - {\ii} \tfrac{1}{2} \Gamma^{(1)}_{f\overline m} } \right|^2  
\nonumber\\[4ex]
=& \; -2 \pi \, \delta(\omega_{fi}) \,
\sum_{\overline m} 
\frac{\left| V_{f\overline m} \, V_{\overline mi} \right|^2}%
{\omega_{\overline mi}^2 + \left(\tfrac{1}{2} \Gamma^{(1)}_{f\overline m}\right)^2 } 
\nonumber\\[4ex]
=& \; -2 \pi \, \delta(\omega_{fi}) \,
\sum_{\overline m}  \frac{2}{\Gamma^{(1)}_{f\overline m}} \,
\frac{\tfrac12 \Gamma^{(1)}_{f\overline m} 
\left| V_{f\overline m} \, V_{\overline mi} \right|^2}%
{\omega_{\overline mi}^2 + \left(\tfrac{1}{2} \Gamma^{(1)}_{f\overline m} \right)^2 } 
\nonumber\\[4ex]
=& \; -4 \pi^2 \, \delta(\omega_{fi}) \,
\sum_{\overline m}  \frac{1}{\Gamma^{(1)}_{f\overline m}} \,
\left| V_{f\overline m} \, V_{\overline mi} \right|^2 \, \delta(\omega_{\overline mi})
\nonumber\\[4ex]
=& \; -4 \pi^2 \, \delta(\omega_{fi}) \,
\sum_{\overline m} \frac{\left| V_{f\overline m} \right|^2 \,
\left| V_{\overline mi} \right|^2}{\Gamma^{(1)}_{f\overline m}}
\delta(\omega_{f\overline m}) 
\nonumber\\[4ex]
=& \; -2\pi \delta(\omega_{ \overline  mi}) \,
\sum_{\overline m} \left| V_{ \overline  mi} \right|^2 
= -\sum_{\overline m} \Gamma^{(1)}_{\overline  mi} \,.
\end{align}
In going from the fourth to the fifth line of the above equation, we have
neglected interference terms. This deserves some comments, which we give by way
of example.  Let us consider a situation with an initial $4S$ state without any
photons, and resonant $2P$ and $3P$ virtual states (each endowed with a single
resonant photon), and a $1S$ final state (with two resonant photons). A
conceivable $2P$--$3P$ interference term would necessitate the final states
$|f\rangle$ to be equivalent in regards to both their atomic components as well
as electromagnetic-field components. 
However, because the emitted resonant photons for
$4S \to 3P \to 1S$ have different energy as compared to $4S \to 2P \to 1S$, the
interference term vanishes. 

The derivation~\eqref{S} clarifies that the subtraction term is nothing but the 
sum of the one-quantum decay rates of the initial state
to all accessible resonant intermediate states.
The result coincides with the lower-order subtraction
term found in Ref.~\cite{Je2009} for the two-photon decay
rate, but the above derivation is much more general.
It means that under this regularization, the 
two-quantum correction to the decay rate is obtained as
\begin{align}
\label{two-quantum-reg}
\overline{\Gamma^{(2)}_{fi}} =& \;
\left. \left( \frac{\rm d}{{\rm d} t} 
\overline{|c_f|^2} \right) \right|_{t=0} 
\nonumber\\
=& \;
2 \pi \, \delta(\omega_{fi}) \,
\left| {\sum_m}  \frac{V_{fm} \, V_{mi}}%
{\omega_{mi} - \frac{1}{2} {\ii} \Gamma^{(1)}_{fm}} \right|^2 -
\sum_{\overline m} \Gamma^{(1)}_{\overline mi} \,,
\end{align}
where we introduce the overlining in order to differentiate
$\overline{\Gamma^{(2)}_{fi}}$ from $\Gamma^{(2)}_{fi}$.
The result~\eqref{two-quantum-reg} is well defined and gauge invariant~\cite{Je2009}.
We also note that 
the total (one-quantum plus two-quantum) 
decay rate of level $|i\rangle$ thus is
\begin{align}
\label{complete}
\Gamma_{i} =& \; 
\sum_{\overline m} \Gamma^{(1)}_{\overline mi} +
\overline{\Gamma^{(2)}_{fi} }
\nonumber\\
=& \; 2 \pi \, \delta(\omega_{fi}) \,
\left| {\sum_m}  \frac{V_{fm} \, V_{mi}}%
{\omega_{mi} - \frac{1}{2} {\ii} \Gamma^{(1)}_{fm}} \right|^2 \,.
\end{align}
This result states, in general terms, that the expression for the 
two-quantum decay rate, in the presence of allowed cascade 
transitions and with propagator denominators regularized
by the total one-quantum decay rate, has to be interpreted as 
a one$+$two quantum decay rate. 

%
%
\subsection{Specialization to an atomic transition}
\label{sec4B}

In view of the result given in 
Eq.~\eqref{known}, it is immediately clear how to apply
Eq.~\eqref{complete} to two-photon transitions in atoms.  Namely, when
Eq.~\eqref{complete} is evaluated for two-photon transitions, the correct
result is obtained when the virtual-state energies in formula~\eqref{known} are
regularized by their total one-photon decay widths [see Eq.~(1) of
Ref.~\cite{Je2009} for a concrete example and extensive
further discussion in
Refs.~\cite{Je2009,ChSu2008,SwHi2008,AmEtAl2009}].  For an atomic two-photon
transition, $\overline{\Gamma^{(2)}_{fi}}$ as written in Eq.~\eqref{two-quantum-reg}
coincides with the imaginary part of the two-loop self-energy due to cut
diagrams with two-photon emission~\cite{Je2007,Je2009}, with the photons
fulfilling the two-photon resonance condition $k_1 +  k_2 = E_{f_A} - E_{i_A}$.

It has been shown in Refs.~\cite{Je2007,Je2009} that 
$\overline{\Gamma^{(2)}_{fi}}$ is of
the order of $\alpha^2 (Z\alpha)^6$ in units of the electron rest mass energy,
and is thus of the same order as the result obtained for two-photon transitions
without cascades~\cite{GM1931}.  It is therefore appropriate to refer to
$\overline{\Gamma^{(2)}_{fi}}$ as a two-photon {\em correction} to the decay rate of an
initial state which otherwise decays via one-photon decay.  For completeness,
we note that the two terms on the right-hand side of Eq.~\eqref{two-quantum-reg}
are both of order $\alpha^2 (Z\alpha)^4$, but their difference is of order
$\alpha^2 (Z\alpha)^6$ and thus smaller by two orders of $Z\alpha$.

The only calculation remaining concerns the verification
of the fact that Eq.~\eqref{expected} reproduces the 
product of one-photon decay rates for the cascade
process $|i_A\rangle \to | m_A \rangle \to | f_A\rangle$.
We use Eq.~\eqref{two-gamma} in order to sum over
the two-photon final states and Eq.~\eqref{continuum}
in order to proceed to the continuum limit.
Summing the cascade term~\eqref{expected} over the 
degrees of freedom of the emitted photons, we obtain
\begin{align}
\label{Cfaia}
& C_{f_A i_A} =  
\left( \frac12 \sum_{\vec{k}_1 \lambda_1} 
\sum_{\vec{k}_2 \lambda_2} \right) \,
\sum_{\overline m} C_{fi}
\\
& =  \frac12 \sum_{\vec{k}_1 \lambda_1} 
\sum_{\vec{k}_2 \lambda_2} \,
\sum_{\overline m} 
\Gamma^{(1)}_{f\overline m} \, 
\Gamma^{(1)}_{\overline mi} 
\nonumber\\
& = 
\frac12
\sum_{\overline m_A}
\sum_{\vec{k}_1 \lambda_1} 
\sum_{\vec{k}_2 \lambda_2} 
\sum_{\vec{k} \lambda} 
2 \pi \delta( E_{f_A} + k_1 + k_2 - E_{{\overline m}_A} - k ) \,
\nonumber\\
& \; \times 2 \pi \delta( E_{{\overline m}_A} - E_{i_A} - k ) \,
\nonumber\\
& \; \times 
\left| \left< {\overline m}_A, 1_{\vec{k}_1 \lambda_1},
1_{\vec{k}_2 \lambda_2} \left|
\left( -e \, \vec{x} \cdot \vec{E} \right) 
\right| i_A, 1_{\vec{k} \lambda}  \right> \right|^2
\nonumber\\
& \; \times
\left| \left< {\overline m}_A, 1_{\vec{k} \lambda} \left|
\left( -e \, \vec{x} \cdot \vec{E} \right) 
\right| i_A, 0 \right> \right|^2 \,.
\nonumber
\end{align}
The summation over $\vec{k}\lambda$ is over both 
polarizations $\lambda$ and over an
energy interval for $k = |\vec{k}|$ which contains the 
resonance frequency of the
intermediate atomic resonant state $| \overline m_A \rangle$.
After performing the sum over $\vec{k}\lambda$
and going to the continuum limit with the help 
of Eq.~\eqref{continuum}, we obtain
\begin{align}
\label{familiar}
& C_{f_A i_A} = 
e^4 \,
\sum_{\overline m_A}
\sum_{\vec{k}_1 \lambda_1} 
\sum_{\vec{k}_2 \lambda_2} 
(2 \pi)^2 \,
\delta( k_1 - ( E_{f_A} - E_{{\overline m}_A}) ) 
\nonumber\\
& \times \delta( k_2 - ( E_{{\overline m}_A} - E_{i_A}) )
\frac{ k_1 }{2 V} 
\left| \left< f_A \left|
\hat{\epsilon}_{\vec{k}_1\lambda_1} \cdot \vec{x} 
\right| {\overline m}_A  \right> \right|^2
\nonumber\\
& \times \frac{ k_2 } {2 V} \, 
\left| \left< {\overline m}_A \left|
\hat{\epsilon}_{\vec{k}_2 \lambda_2} \cdot \vec{x} 
\right| i_A \right> \right|^2 
\nonumber\\
= & \; 
\sum_{\overline m_A}
\left( \frac{4 \alpha}{3} \, ( E_{f_A} - E_{{\overline m}_A})^3
\left| \left< f_A \left| \vec{x} \right| {\overline m}_A  \right> \right|^2 \right. \,
\nonumber\\
& \times
\left. \frac{4 \alpha}{3} \, ( E_{{\overline m}_A} - E_{i_A} )^3
\left| \left< {\overline m}_A \left| \vec{x} \right| i_A  \right> \right|^2 \right) 
\nonumber\\
= & \;
\sum_{\overline m_A}
\Gamma^{(1\gamma)}_{f_A {\overline m}_A}
\Gamma^{(1\gamma)}_{{\overline m}_A i_A} \,.
\end{align}
This result confirms that the cascade terms can indeed be written
as the product of atomic one-photon decay rates.

%
%
\section{Conclusions}
\label{sec5}

In this paper, we have reviewed the formulation of a transition 
with a single quantum jump within time-dependent perturbation theory
(see Sec.~\ref{sec2A}). The result, which is Fermi's golden rule 
[see Eq.~\eqref{FGR}], is evaluated for an atomic dipole transition
in Sec.~\ref{sec2B}. We find that the familiar result for the 
one-photon decay rate [see Eq.~\eqref{manif}] is obtained after a
summation/integration over the degrees of freedom of the emitted
photon in the continuum limit, as given in Eq.~\eqref{continuum}.
The general formulation of a transition with two quantum jumps 
is carried out in Sec.~\ref{sec3A}, within second-order time-dependent
perturbation theory. The result for the two-quantum decay 
rate $\Gamma^{(2)}_{fi}$ as given in Eq.~\eqref{Gamma2without}
is valid if there are no resonant intermediate states through which 
a cascade decay could possibly proceed. The specialization to a 
transition with two quantum jumps is carried out in Sec.~\ref{sec3B},
where it is shown that a summation over the 
two-photon final states of the process [see Eq.~\eqref{two-gamma}]
yields the familiar result~\eqref{known} for a two-photon transition
rate in a hydrogenlike ion (such as the $2S$--$1S$ decay).
Cascade contributions are analyzed in Sec.~\ref{sec4A}.

By isolating the resonant states within the intermediate 
states of the process, we obtain the cascade contribution~\eqref{cascade}
after differentiating the probability of finding the system in 
the final state {\em twice} with respect to the elapsed time.
Subtracting the contribution of the resonant states within the 
{\em first} time derivative of said probability, 
we obtain an expression for the two-quantum correction to the 
decay rate of a system which can simultaneously decay 
via cascades and two-quantum transitions.
If we use the identification~\eqref{ident} for the 
regularization parameter $\eta$ that parameterizes the 
width of the intermediate states,
then the effect of subtracting the resonant states from the 
two-quantum decay rate is equivalent to the subtraction of the 
total one-photon decay rate of the initial state 
[see Eq.~\eqref{two-quantum-reg}]. The difference of the 
formal two-quantum decay rate (with the propagator denominators 
of the resonant states regularized by their widths)
and the one-photon decay rate therefore constitutes
the two-quantum correction $\overline{\Gamma^{(2)}_{fi}}$
to the decay rate, as noted in Eq.~\eqref{two-quantum-reg}. 

The formal two-quantum decay rate
(with the propagator denominators
of the resonant states regularized by their widths)
therefore constitutes more than its name might suggest:
namely, according to Eq.~\eqref{complete}, 
it is the sum of the one-quantum decay rate 
and of the two-quantum correction and therefore 
constitutes, in some sense, a one$+$two quanta decay rate.
This finding provides a formal and general justification for the 
observation made in Ref.~\cite{Je2009}: namely, that the 
formal two-photon decay rate of, say, a $4S$ state in a hydrogenlike ion
contains the one-photon decay rate of $4S$ as a lower-order term
which needs to be subtracted in order to obtain the pure
two-photon correction. It is instructive to also remember that 
competing one- and two-photon
decays are not restricted to atomic hydrogen,
but also occur in other atomic systems of fundamental
importance like rubidium~\cite{ToEtAl1997}.

Finally, in Sec.~\ref{sec4B}, we analyze the cascade contribution 
for atomic transitions and show that under the 
appropriate normalization of the two-photon final state
[see Eqs.~\eqref{two-gamma} and~\eqref{Cfaia}], the familiar
result is obtained [see Eq.~\eqref{familiar}].
The results reported here have an interesting consequence,
because they imply that the one$+$two-photon decay 
rate~\eqref{complete} 
gives the correct differential-in-energy rate coefficients for recombination
codes~\cite{ChSu2008,SwHi2008}, where the spectrum of emitted photons 
from both two-photon and cascade transitions
is needed over the entire range of resonant and nonresonant frequencies.
In particular,
the corresponding expression~\eqref{complete} can be used even in the
vicinity of resonant bound-state poles, provided these are regularized by their
respective decay widths.

One particular limitation of the treatment discussed here is immediately
obvious. Adiabatic perturbation theory does not make any statement about the
time points $t'$ and $t''$ where the two photons are emitted. Thus, the total
regularized two-photon decay rate as described by Eq.~\eqref{complete} contains
both correlated emission (with a small time difference $|t' - t''|$) and also
sequential emission (with a large time difference $|t' - t''|$).  Since it is
only the photon spectrum, not the correlation~\cite{RaSuFr2008}, 
that matters for cosmological
recombination, we can use the regularized two-photon decay
rate~\eqref{complete}, differential in energy, with good accuracy for
recombination codes~\cite{ChSu2008,SwHi2008} which need the
differential-in-energy spectrum of the emitted photons as input. In particular,
this means that it is not necessary to distinguish specific contributions to
the complete rate~\eqref{complete}; this rate contains both cascade photons and
correlated two-photon processes in a natural way.  Therefore, it is
understandable that the authors of Ref.~\cite{LaSoPl2009} could not give a
unique value to their parameter $\Delta\omega$; this parameter was introduced
in order to distinguish between cascade photons and two-photon decays.
If one would like to make a more refined distinction between cascade
photons and correlated two-photon processes, then one has to go beyond
adiabatic perturbation theory and analyze the dependence of the process on the
emission times $t'$ and $t''$, including loss of correlation as a function of
$|t' - t''|$, which may be process- as well as environment-dependent (e.g.,
there may be a dependence on the average mean free path of the atoms in their
environment). 

Our article illustrates both the usefulness but also the limitations of
adiabatic perturbation theory.  Namely, if we apply the
regularization~\eqref{ident} consistently, to both the first term as well as
the second term on the right-hand side of~\eqref{eq3}, then the subtraction
term $S$ defined in Eq.~\eqref{defS} attains a finite value and can be
evaluated in closed form [see Eq.~\eqref{S}].  Furthermore, as shown in
Ref.~\cite{Je2009}, if the regularization~\eqref{ident} is applied to a two-photon
transition in atoms, then there are significant cancellations between the two
terms on the right-hand side of~\eqref{two-quantum-reg}, which are both of order
$\alpha^2(Z\alpha)^4$, but their difference is of order $\alpha^2(Z\alpha)^6$,
where $Z$ is the nuclear charge number, and $\alpha$ is the fine-structure
constant.  Two-photon decay rates are of order $\alpha^2(Z\alpha)^6$.  As
evident from Eq. (11) of Ref.~\cite{Je2009} and from Eq.~\eqref{S} in the current
work, the cancellation of the lower-order terms depends on the particular
choice of the regularization. Within adiabatic perturbation theory, the
regularization~\eqref{ident} thus appears to be the only one which leads to a consistent
removal of the infinities that plague the two-quantum decay rate in the
presence of allowed cascade transitions. Therefore, our article offers---for the 
first time in the literature, to the best of our knowledge---a connection 
of the adiabatic parameter $\eta$ used in  time-dependent perturbation
theory to a physical concept, namely, the lifetime of virtual intermediate states.

An interesting connection to the theory of energy shifts of atomic levels can
be drawn. Low~\cite{LaSoPlSo2001,JeEvKePa2002,Lo1952} observed that the calculation of
energy shifts of excited states of hydrogenlike ions becomes problematic at
order $\alpha^2(Z\alpha)^6$, due to interference effects of the resonance line
shapes of atomic levels of different principal quantum number. It has been
argued that at order $\alpha^2(Z\alpha)^6$, two-loop energy shifts of excited
states cannot be uniquely associated any more with a particular atomic level,
due to the predictive limits of adiabatic perturbation
theory~\cite{LaSoPlSo2001,JeEvKePa2002,Lo1952}.  The decay rate at order
$\alpha^2(Z\alpha)^6$ constitutes the imaginary part of the energy shift of
that same order. It is thus not surprising that its calculation requires
considerable effort within the formalism of adiabatic perturbation theory.

%
%
\section*{ACKNOWLEDGMENTS}

The author acknowledges helpful conversations with Professor A.~I.~Milstein.
Helpful remarks of an anonymous referee are also gratefully acknowledged.
This research was supported by the National Science Foundation (PHY--8555454)
and by a precision measurement grant from the National Institute of Standards
and Technology.

\end{document}